\documentclass[12pt]{article}
\usepackage{epsfig}
\usepackage{graphicx}
\usepackage{amsfonts}
\usepackage{bbm}
\newcommand\C{{\mathbbm{C}}}

\newcommand\Pp{{\mathbbm{P}}}
\newcommand\R{{\mathbbm{R}}}

\begin{document}
\thispagestyle{empty}

\begin{center}\begin{Large}\begin{bf}
The geometry of one-loop amplitudes\\
\end{bf}\end{Large}\vspace{.75cm}
\vspace{0.5cm} Oliver Schnetz\footnote{schnetz@mi.uni-erlangen.de}\\
Department Mathematik,\\
Universit{\"a}t Erlangen-N{\"u}rnberg, Germany\\
October 2010\\
\vspace{1cm}\baselineskip=35pt
\end{center}
 
\begin{abstract}
We review a reduction formula by Petersson that reduces the calculation of a one-loop amplitude with $N$ external lines in $n<N$ space-time dimensions
to the case $n=N$ and give it a geometric interpretation. In the case $n=N$ the calculation of the euclidean amplitude is shown to be equivalent to
the calculation of the volume of a tetrahedron spanned by the momenta in ($n-1$)-dimensional hyperbolic space. The underlying geometry is intimately
linked to the geometry of the reduction formula.
\end{abstract}
\tableofcontents

\section{Introduction}
The last decade has seen a renewed interest in perturbative quantum field theory (pQFT).
On the one hand, progress has been achieved on amplitudes with many legs and a low number of loops (zero or one)
\cite{W1}, \cite{W2}, \cite{B1} (and the references therein).
From an experimentalist point of view these results are vital for the analysis of, e.g., the LHC-data.
On the other hand the study of many loops with a low number of external legs is important for the
understanding of high precision experiments like the measurement of the anomalous magnetic moment of
the electron \cite{O1}, \cite{G1}, \cite{H1}.

This article focuses on the first aspect of pQFT. Because of their immediate relevance for physical predictions one-loop amplitudes have attracted much attention.
In the 1960s it was observed that the complexity of the amplitude saturates when the number of external particles exceeds the dimension
of space-time \cite{BR2}, \cite{HA}, \cite{KT}. In 1965 Petersson \cite{PE} found the residue-type reduction formula Eq.\ (\ref{30}),
$$\int{\rm d}^np\frac{1}{Q_1\cdots Q_N}={\rm Re}\sum_{I\subset \{1,...,N\}\atop |I|=n}
\frac{1}{\prod\limits_{i\in\!\!\!\!/\,I}Q_i}\Bigg|_{p:{Q_i(p)=0\atop i\in I}}
\int{\rm d}^np\,\frac{1}{\prod\limits_{i\in I}Q_i}$$
which reduces the number of propagators $Q_i$ in the integral to the space-time dimension $n$ at the expense of a pre-factor which is
a rational function in the momenta and masses of the particles.

At the same time it was observed that in space-time dimensions two and three the amplitude is given by logarithms of algebraic
functions in the momenta and masses \cite{KT}, \cite{NI}. In space-time dimension four (and five) one needs dilogarithms to express the amplitude \cite{WU}
(for a modern mathematical account see \cite{BK1}).
Only in 1998 it was observed by Davydychev and Delbourgo that the simplicity of the results can be understood in a geometrical language:
The amplitudes are volumes of tetrahedra in---depending on the masses and momenta---spherical or hyperbolic geometries \cite{DA}.
In this article we show that one obtains a unified hyperbolic picture if one works in euclidean instead of Lorentzean space-time.
For $n$ external particles in $n$ space-time dimensions one obtains Eq.\ (\ref{46}),
$$\int{\rm d}^np\frac{1}{Q_1\cdots Q_n}=\frac{(2\pi)^\frac{n}{2}\Gamma(\frac{n}{2})\hbox{vol}_{H^{n-1}}(\Sigma)}
{[(-1)^{n-1}\hbox{det}((p_i-p_j)^2+m_i^2+m_j^2)_{i,j}]^{1/2}}.$$
where the hyperbolic tetrahedron $\Sigma$ has the following geometric interpretation: By momentum conservation the external momenta sum up to zero.
Their vectors span an ($n-1$)-dimensional tetrahedron (in three dimensions they are the sides of a triangle (see Fig.\ 3), in four dimensions
they form a cycle of four edges that spans a tetrahedron). This tetrahedron is $\Sigma$ in the projective hyperbolic model (where geodesics are straight lines).
The location of the 'sphere at infinity' is determined by the masses of the internal particles.
If the particles are massless the sphere at infinity is spanned by the vertices of the (then ideal) tetrahedron.
With this picture one can rephrase the above formula entirely in terms of geometrical quantities, basically as the ratio of the hyperbolic
over the euclidean volume of $\Sigma$, Eq.\ (\ref{52}),
$$\int{\rm d}^np\frac{1}{Q_1\cdots Q_n}=\frac{\hbox{vol}(S^n_{1/2})\hbox{vol}_{H^{n-1}}(\Sigma)}{r\,\hbox{vol}_{\R^{n-1}}(\Sigma)}$$
where $r$ is the euclidean radius of the sphere at infinity and vol$(S^n_{1/2})$ is the volume of the $n$-dimensional euclidean sphere of radius $1/2$.
Remarkably the volume of any hyperbolic tetrahedron in odd-dimensional hyperbolic space (corresponding to even-dimensional space-time) can be
expressed in terms of a (half-)sum of volumes of ideal tetrahedra (see Sect.\ \ref{step5} for a precise statement). Because ideal tetrahedra correspond to massless particles any
one-loop amplitude of massive particles in even-dimensional space-time can be expressed in terms of one-loop amplitudes of massless particles (whose external momenta are algebraic
functions of the masses and momenta of the original particles).

The aim of this article is to present the above results in a unified geometric framework. The necessary algebra is organized in 5 elementary steps which
are fully explained and elaborated. We end this article by the short example of a massive two-loop amplitude in two dimensions which shows that beyond the
one-loop level polylogarithms are in general replaced by elliptic or even more complicated polylogarithms. A hyperbolic interpretation of general
two loop amplitudes is hence not possible.

\subsection{A four-dimensional example}\label{4dex}

A particularly interesting situation is the one-loop amplitude of a four dimensional massless theory. In this case one can use quaternions to represent the
momenta which gives an interesting shortcut to hyperbolic geometry. We start from the amplitude (see Fig.\ 1)
\begin{equation}\label{0a}
A(p_1,p_2,p_3,p_4)=\int_{\R^4}{\rm d}^4p\frac{1}{|p-p_1|^2|p-p_2|^2|p-p_3|^2|p-p_4|^2}.
\end{equation}
Now, the transformation $p\mapsto p^{-1}+p_3$ leads to
\begin{equation}\label{0a1}
A(p_1,p_2,p_3,p_4)=\int_{\R^4}{\rm d}^4p\frac{1}{|1+p\,p_{1,3}|^2|1+p\,p_{2,3}|^2|1-p\,p_{3,4}|^2},
\end{equation}
where $p_{i,j}=p_j-p_i$. Next, $p\mapsto p-p_{1,3}^{-1}$ followed by $p\mapsto -p_{1,3}^{-1}\,p\,p_{4,1}\,p_{3,4}^{-1}$ gives
\begin{eqnarray}\label{0b}
A(p_1,p_2,p_3,p_4)&=&\frac{1}{|p_{2,3}|^2|p_{4,1}|^2}\int_{\R^4}{\rm d}^4p\frac{1}{|p|^2|p-1|^2|p-q|^2},\\
&&\hbox{with}\quad q=p_{1,2}\,p_{2,3}^{-1}\,p_{3,4}\,p_{4,1}^{-1}.
\end{eqnarray}
To solve the integral we first assume $|q|<1$ and expand the propagators into series of Gegenbauer polynomials $C_n$ (see \cite{C4})
\begin{equation}\label{0c}
\frac{1}{|p-q|^2}=\frac{1}{|p||q|}\sum_{n=0}^\infty C_n(p,q)\left(\frac{|p|}{|q|}\right)_<^{n+1},\hbox{ where }x_<=\hbox{min}(x,1/x).
\end{equation}
By the orthogonality of the Gegenbauer polynomials ($p=|p|\,\hat{p}$)
\begin{equation}\label{0d}
\int_{S^3}\frac{{\rm d}\hat{p}}{2\pi^2}\,C_n(p,q)C_m(p,r)=\frac{\delta_{n,m}}{n+1}C_n(q,r)
\end{equation}
we obtain 
\begin{equation}\label{0e}
I(q)=\int_{\R^4}{\rm d}^4p\frac{1}{|p|^2|p-1|^2|p-q|^2}=2\pi^2\sum_{n=0}^\infty\left(\frac{1}{(n+1)^2}-\frac{\ln|q|}{n+1}\right)C_n(1,q)|q|^n.
\end{equation}
With the explicit expression for the Gegenbauer polynomials we have
\begin{equation}\label{0e1}
C_n(1,q)|q|^n=\frac{q^{n+1}-\bar{q}^{n+1}}{q-\bar{q}}
\end{equation}
yielding, with Li$_2(q)=\sum_{n=1}^\infty q^n/n^2$, (note that the sign ambiguity in the quaternionic imaginary part cancels)
\begin{equation}\label{0h}
I(q)=2\pi^2\frac{\rm{Im\;[Li}_2(q)+\ln|q|\ln(1-q)]}{\rm{Im}\;q}.
\end{equation}
The numerator is the Bloch-Wigner dilogarithm, (\cite{BL}, \cite{TH}, \cite{ZA})
\begin{eqnarray}\label{0f}
D(x)&=&{\rm Im}\,[\hbox{Li}_2(x)+\ln|x|\ln(1-x)]\quad\hbox{ for }|x|<1,\nonumber\\
D(x)&=&-D(1/x)\hspace{37mm}\hbox{ for }|x|>1.
\end{eqnarray}
The above transformation law for $x\mapsto 1/x$ allows us to write the result for all $q$ (as can be seen
by a transformation $p\mapsto 1/p$ in $I(q)$). For the amplitude of the one-loop graph we finally obtain
\begin{equation}\label{0g}
A(p_1,p_2,p_3,p_4)=(2\pi)^2\frac{D(p_{1,2}\,p_{2,3}^{-1}\,p_{3,4}\,p_{4,1}^{-1})}{2\,\rm{Im}\;p_{1,2}\,\overline{p_{2,3}}\,p_{3,4}\,\overline{p_{4,1}}}
\end{equation}
which is the quaternionic version of Eq.\ (\ref{46}). If the four external momenta $p_{1,2}$, $p_{2,3}$, $p_{3,4}$, $p_{4,1}$ fall into a plane
we can interpret this plane as the complex number field embedded into the quaternions. This makes the momenta complex and the
Bloch-Wigner dilogarithm is known to give the (signed) volume of the ideal hyperbolic tetrahedron which is spanned by the momenta $p_1$, $p_2$, $p_3$, $p_4$ in the half-space
model of hyperbolic geometry. In this model the sphere at infinity degenerates to the complex plane spanned by the momenta.
The validity of Eq.\ (\ref{46}) implies that the hyperbolic interpretation still holds for any quaternionic momenta. In the general case the
momenta $p_1$, $p_2$, $p_3$, $p_4$ span a two-dimensional sphere which defines the sphere at infinity of a three-dimensional hyperbolic space.
In the projective model the edges of the tetrahedron are the straight lines $p_{i,j}$ from $p_i$ to $p_j$.
Hence momentum conservation $p_{1,2}+p_{2,3}+p_{3,4}+p_{4,1}=0$ provides an edge-cycle that spans the tetrahedron (see Fig.\ 5 for the three-dimensional massive case).

{\bf Acknowledgements.}
I am grateful to Ruth Kellerhals and Andreas Bernig for helpful discussions on hyperbolic tetrahedra.

\section{The one-loop amplitude}
In this section we reduce the computation of a one-loop amplitude in several steps.
We consider a bosonic QFT in $n$ space-time dimensions where each propagator has its own mass.
We choose an euclidean metric which is best suited to exhibit a universal geometrical picture (see \cite{DA} for
a Minkowskian version). The one-loop amplitude with $N$ external legs is an integral over a product of $N$ propagators (see Fig.\ 1).

\begin{figure}[ht]
\hspace{5cm}\epsfig{file=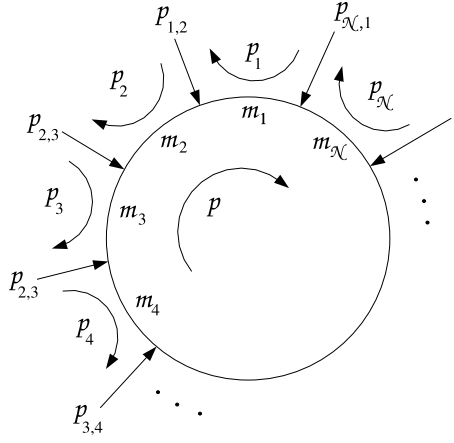,width=6cm}
\caption{Feynman diagram of the one-loop amplitude. If the theory has four-valent vertices the external momentum $p_{i,i+1}$ is the sum of
the ingoing momenta at the corresponding vertex.}
\end{figure}

We define
\begin{equation}\label{1}
Q_i=(p-p_i)^2+m_i^2,\quad i=1,\dots ,N.
\end{equation}
Here $p,p_i$ are vectors in $\R^n$ endowed with the euclidean scalar product $p^2=(p^{(1)})^2+\dots+(p^{(n)})^2$ (we indicate
vector-components by upper braced indices).
The ingoing external momenta are
\begin{equation}\label{1a}
p_{i,i+1}=p_{i+1}-p_i
\end{equation}
where $p_{N+1}=p_1$ and $p_{N,N+1}=p_{N,1}$. By notation we have implemented the conservation of the total momentum
$\sum_{i=1}^Np_{i,i+1}=0$. In the following we assume that the ingoing momenta have general values and do not consider degenerate situations.
The amplitude is
\begin{equation}\label{2}
A(p_1,m_1,\dots,p_N,m_N)=\int_{\R^n}{\rm d}^np\frac{1}{Q_1\cdots Q_N}.
\end{equation}
It is useful to give each propagator its own mass in order to exhibit the geometrical role of the masses.
By a shift in the integration variable one may set one of the momenta to zero, e.g.\ $p_1=0$.
Here, we prefer to keep the symmetry under permutations of the $p_i, m_i$.

We reduce the calculation of the amplitude in the following steps:

\begin{enumerate}\setcounter{enumi}{-1}
\item All but one propagators are massless (optional),
\item $N\leq n+1$,
\item $N\leq n$,
\item Calculation of volumes of hyperbolic tetrahedra in $n-1$ dimensions,
\item $n$ even,
\item All propagators are massless, calculation of volumes of ideal hyperbolic tetrahedra in odd dimensions.
\end{enumerate}

Step 0 is not needed in the subsequent steps. It is still of interest because it can be helpful in
more complicated situations (like two or more loops). Moreover, in step 0 we motivate the geometrical concepts of the later steps.

\subsection{Step 0: $m=0$ for all but one propagator}\label{step0}
We define a massless propagator $1/Q_0$ with
\begin{equation}\label{3}
Q_0=(p-p_0)^2
\end{equation}
and find by a partial fraction decomposition
\begin{equation}\label{3a}
\frac{1}{Q_1Q_2}=\frac{\lambda}{Q_0Q_1}+\frac{1-\lambda}{Q_0Q_2}
\end{equation}
if
\begin{eqnarray}\label{4}
p_0&=&(1-\lambda)p_1+\lambda p_2,\nonumber\\
\lambda&=&\frac{p_{1,2}^2-m_1^2+m_2^2\pm\sqrt{[p_{1,2}^2+(m_1+m_2)^2][p_{1,2}^2+(m_1-m_2)^2]}}{2p_{1,2}^2}.
\end{eqnarray}
This slightly awkward looking formula has a geometrical interpretation.
First, we notice that we have two choices $p_0^\pm$ for $p_0$ according to the $\pm$ sign in (\ref{4}).
Second, we find that $p_0^\pm$ is collinear to $p_1$ and $p_2$.
We thus may represent $p_0^\pm$, $p_1$, $p_2$ as points on a line and $p_0^\pm$ are
found by introducing a 'mass-axis' and plotting the four points $(p_1,\pm m_1)$, $(p_2,\pm m_2)$ (see Fig.\ 2).
The points $p_0^\pm$ are recovered as the intersections of a circle through the four points with the $p$-axis:
The intersections have the coordinates $(p_0^\pm,0)$ with $p_0^\pm$ as in Eq.\ (\ref{4}) and the second
coordinate 0 rendering $p_0^\pm$ 'massless'.

\begin{figure}[ht]
\hspace{4cm}\epsfig{file=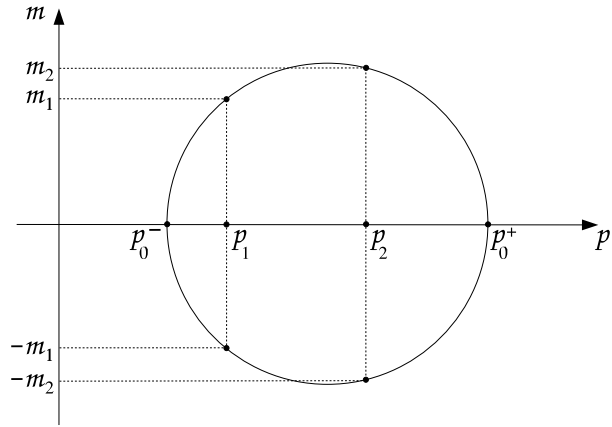,width=8cm}
\caption{Decomposition into massless propagators.}
\end{figure}

We may anticipate two results from the above considerations: First, it will turn out to be useful to introduce an extra
mass coordinate 'vertical' to the momenta. Second, the advent of (half-)circles hints to hyperbolic geometry (think
of the Poincar\'e half-space model of $H^{n+1}$).
We will see later (Step 5, Sect.\ \ref{step5}) that the decomposition in Eq.\ (\ref{3a}) amounts to a decomposition of
hyperbolic tetrahedra into tetrahedra with vertices at infinity.

Repeated application of Eq.\ (\ref{3a}) allows us to decompose the one-loop amplitude into amplitudes
with all but one propagator massless. In Sect.\ \ref{step5} we will see by purely geometrical means that it is possible
to find a decomposition with {\em all} propagators massless: In odd-dimensional hyperbolic space we can decompose any tetrahedron
into ideal tetrahedra (with vertices at infinity). The new momenta are algebraic functions of the original momenta and masses similar to Eq.\ (\ref{4}).
This means in the physical context that we can reduce the calculation
of massive one-loop amplitudes to the calculation of amplitudes with massless propagators.

We will not make use of Eq.\ (\ref{3a}) in the following steps but rather keep the full symmetry.

\subsection{Step 1: $N\leq n+1$}\label{step1}
To achieve the reduction from $N>n+1$ to $N=n+1$ it is sufficient to find a partial fraction decomposition of $n+2$ propagators into
a sum of products of $n+1$ propagators. Repeated use of this relation reduces the number of propagators
until $N=n+1$.

To find this decomposition we need $n+2$ coefficients $\lambda_i$ which do not depend on $p$ and
fulfill the identity
\begin{equation}\label{5}
1=\lambda_1Q_1+\lambda_2Q_2+\dots+\lambda_{n+2}Q_{n+2}.
\end{equation}
Upon dividing Eq.\ (\ref{5}) by $Q_1Q_2\cdots Q_{n+2}$ we obtain the desired relation.
Solving Eq.\ (\ref{5}) for the $\lambda_i$ means to match the coefficients in front of $p^2$, $p^{(1)}$, \dots, $p^{(n)}$, and the constant 1.
The linear system has a solution if the determinant of its coefficient matrix is non-zero:
\begin{equation}\label{6}
\hbox{det}\left(\begin{array}{cccc}
1&1&\cdots&1\\
-2p_1^{(1)}&-2p_2^{(1)}&\cdots&-2p_{n+2}^{(1)}\\
\vdots&\vdots&&\vdots\\
-2p_1^{(n)}&-2p_2^{(n)}&\cdots&-2p_{n+2}^{(n)}\\
p_1^2+m_1^2&p_2^2+m_2^2&\cdots&p_{n+2}^2+m_{n+2}^2\end{array}\right)\neq 0.
\end{equation}
It is easy to see that for generic values of $p_1$, $m_1$, \dots $p_{n+2}$, $m_{n+2}$ the required condition holds.

Here, it is sufficient to know that the decomposition exists. We are not interested in the precise values
for the $\lambda_i$ because in the next step we continue the decomposition to $N=n$ and obtain an expression for
the coefficients as a side product.

\subsection{Step 2: $N\leq n$}
If we try to continue the decomposition of Step 1 to $N=n$ we find that we have $n+1$ variables $\lambda_i$ to fulfill a system of
$n+2$ equations. In general this system will have no solution. However, we actually need a decomposition that is valid {\em after} the
integration over $p$. We are free to add terms that vanish when integrated over $p$. We will see that this gives us an extra propagator $1/Q$ to our disposal.
With this propagator we can proceed as in Eq.\ (\ref{5}) and solve
\begin{equation}\label{7}
1=\lambda_1Q_1+\lambda_2Q_2+\dots+\lambda_{n+1}Q_{n+1}+\lambda_{n+2}Q.
\end{equation}
for $p$-independent $\lambda_i$. Division of Eq.\ (\ref{7}) by $Q_1Q_2\cdots Q_{n+1}$ gives the desired decomposition provided
\begin{equation}\label{8}
\int{\rm d}^np\frac{Q}{Q_1Q_2\cdots Q_{n+1}}\stackrel{!}{=}0.
\end{equation}
The construction of $Q$ starts with the observation that the vector-valued integral (convergent for $n\geq 2$)
\begin{equation}\label{9}
I_i=\int{\rm d}^np\frac{p_i-p}{Q_1Q_2\cdots Q_n}
\end{equation}
depends (by a shift $p\to p+p_i$) on the vectors $p_j-p_i$ ($j\neq i$) only. By symmetry it has to assumes its value in the $(n-1)$-dimensional
subspace spanned by these vectors. This subspace is parallel to the hypersurface spanned by the tips of the vectors $p_1$, \dots, $p_n$
and it is the same for all $i=1,\dots,n$. Thus the $n$ vectors $I_i$ have to be linearly dependent.
The $n$ by $n$ matrix $M=(I_1,I_2,\dots I_n)$ has zero determinant. We introduce the $n+1$ by $n+1$ determinant 
\begin{equation}\label{10}
\hbox{det}(\bar{p},\bar{p}_1,\dots,\bar{p}_n)\equiv\hbox{det}\left[\left(\!\begin{array}{c}1\\p\end{array}\!\right),
\left(\!\begin{array}{c}1\\p_1\end{array}\!\right),\dots,\left(\!\begin{array}{c}1\\p_n\end{array}\!\right)\right]
\end{equation}
and recall that
\begin{equation}\label{11}
\hbox{det}(\bar{p},\bar{p}_1,\dots,\bar{p}_n)=\hbox{det}(p_1-p,\dots,p_n-p)=\pm \,[\hbox{det}\left((p_i-p)\cdot(p_j-p)\right)_{i,j}]^{1/2}
\end{equation}
gives $\pm n!$ times the volume of the simplex spanned by the tips of the vectors $p,p_1,\dots ,p_n$.
With this notation the linear dependence of the $I_i$ can be expressed as
\begin{equation}\label{12}
\int{\rm d}^np\;\frac{\hbox{det}(\bar{p},\bar{p}_1,\dots,\bar{p}_n)}{Q_1Q_2\cdots Q_n}=0.
\end{equation}
Now we define $Q$ by the determinant of an $n+2$ by $n+2$ matrix (compare Eq.\ (\ref{6})):
\begin{equation}\label{13}
Q=\frac{(-1)^{n+1}}{\hbox{det}(\bar{p}_1,\dots,\bar{p}_{n+1})}\;
\hbox{det}\left(\begin{array}{cccc}
1&1&\cdots&1\\
p^{(1)}&p_1^{(1)}&\cdots&p_{n+1}^{(1)}\\
\vdots&\vdots&&\vdots\\
p^{(n)}&p_1^{(n)}&\cdots&p_{n+1}^{(n)}\\
p^2&p_1^2+m_1^2&\cdots&p_{n+1}^2+m_{n+1}^2\end{array}\right).
\end{equation}
It is readily checked that $Q$ has the form of an inverse propagator and upon adding $p^2$ times the first row and
$-2p^{(i)}$ times the $(i+1)$st row to the last row we see that
\begin{equation}\label{14}
Q=\frac{(-1)^{n+1}}{\hbox{det}(\bar{p}_1,\dots,\bar{p}_{n+1})}\;
\hbox{det}\left(\begin{array}{cccc}
1&1&\cdots&1\\
p^{(1)}&p_1^{(1)}&\cdots&p_{n+1}^{(1)}\\
\vdots&\vdots&&\vdots\\
p^{(n)}&p_1^{(n)}&\cdots&p_{n+1}^{(n)}\\
0&Q_1&\cdots&Q_{n+1}\end{array}\right).
\end{equation}
Developing with respect to the last row we find that
\begin{equation}\label{14a}
Q=\sum_{i=1}^{n+1}\frac{(-1)^iQ_iD_i}{\hbox{det}(\bar{p}_1,\dots,\bar{p}_{n+1})}
\end{equation}
where we used the shorthand
\begin{equation}\label{15}
D_i=\hbox{det}(\bar{p},\bar{p}_1,\dots,\bar{p}_{i-1},\bar{p}_{i+1},\dots,\bar{p}_{n+1}).
\end{equation}
With this $Q$ in Eq.\ (\ref{8}) we obtain a sum of $n+1$ terms each of which is zero by Eq.\ (\ref{12}).

Before we determine the $\lambda_i$ in Eq.\ (\ref{7}) it is worthwhile
to have a closer look at $Q$. It is an inverse propagator thus having the form $Q=(p-q)^2+m^2$ for some pair $(q,m)$.
What is this pair? We develop the determinant in Eq.\ (\ref{13}) with respect to the first column and obtain
\begin{equation}\label{18}
Q=p^2-2p^{(1)}q^{(1)}-\dots-2p^{(n)}q^{(n)}+C
\end{equation}
which defines $q$ and $m^2=C-q^2$. By Cramer's rule we know that the vector $v^{\rm T}=(v^{(0)},\dots,v^{(n)})
=(C,-2q^{\rm T})$ is (for generic $p_i$) the unique solution of the linear system
\begin{equation}\label{19}
\left(\begin{array}{cccc}
1&p_1^{(1)}&\cdots&p_1^{(n)}\\
\vdots&\vdots&&\vdots\\
1&p_{n+1}^{(1)}&\cdots&p_{n+1}^{(n)}
\end{array}\right)\cdot v+
\left(\begin{array}{c}
p_1^2+m_1^2\\
\vdots\\
p_{n+1}^2+m_{n+1}^2
\end{array}\right)=0.
\end{equation}
The first equation gives the relation
\begin{equation}\label{20}
C-2p_1\cdot q+p_1^2+m_1^2=0.
\end{equation}
If we use this to eliminate $C=v^{(0)}$ in Eq.\ (\ref{19}) we obtain
\begin{equation}\label{21}
-2\left(\begin{array}{ccc}
p_2^{(1)}-p_1^{(1)}&\cdots&p_2^{(n)}-p_1^{(n)}\\
\vdots&&\vdots\\
p_{n+1}^{(1)}-p_1^{(1)}&\cdots&p_{n+1}^{(n)}-p_1^{(n)}
\end{array}\right)\cdot q+
\left(\begin{array}{c}
p_2^2+m_2^2-p_1^2-m_1^2\\
\vdots\\
p_{n+1}^2+m_{n+1}^2-p_1^2-m_1^2
\end{array}\right)=0.
\end{equation}
We may use Eq.\ (\ref{21}) to solve for $q$ and thereafter determine $C$ and hence $m$ by Eq.\ (\ref{20}).
The result can be interpreted as a higher-dimensional analog of the situation in Fig.\ 2 (see Fig.\ 3 for $n=2$):
We plot the $2(n+1)$ vectors $(p_i,\pm m_i)$ in $\R^{n+1}$ and draw the sphere through these points. The center
of the sphere $(c,0)$ and its radius $r$ are determined by the equations

\begin{figure}[ht]
\hspace{2cm}\epsfig{file=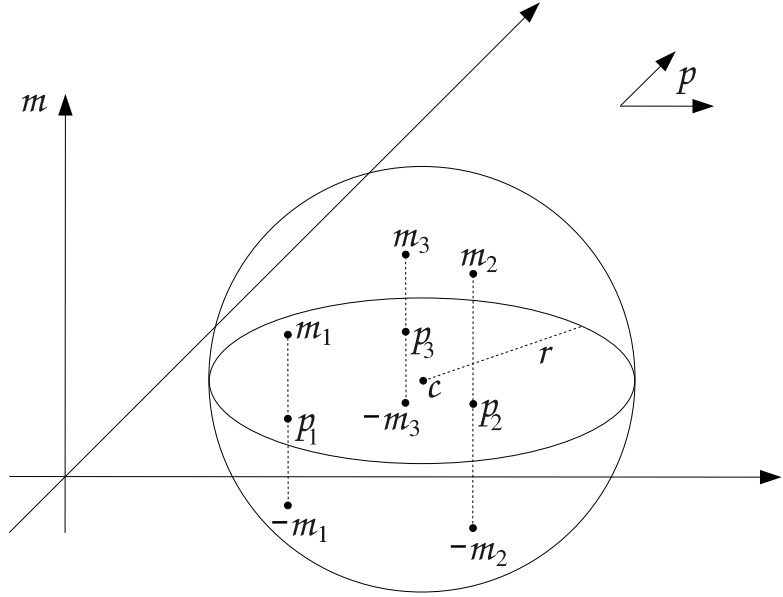,width=12cm}
\caption{The sphere spanned by the momenta and masses.}
\end{figure}

\begin{equation}\label{22}
(p_i-c)^2+m_i^2=r^2,\quad i=1,\dots,n+1.
\end{equation}
If we subtract the first equation ($i=1$) from the others we obtain
\begin{equation}\label{23}
p_i^2-p_1^2-2(p_i-p_1)\cdot c+m_i^2-m_1^2=0,\quad i=2,\dots,n+1.
\end{equation}
with the solution given by Eq.\ (\ref{21}). Hence $c=q$ and by the first equation we have
\begin{equation}\label{24}
p_1^2-2p_1\cdot q+q^2+m_1^2=r^2.
\end{equation}
Comparison with Eq.\ (\ref{20}) and $C=m^2+q^2$ yields $m^2=-r^2$. The propagator $1/Q=1/[(p-q)^2+m^2]$ is thus 'tachyonic' and we find
that $q$ is the center of the sphere in $\R^{n+1}$ and $m$ is its imaginary radius,
\begin{equation}\label{25}
(q,m)=(c,{\rm i} r).
\end{equation}

Now, we want to determine the coefficients of $\lambda_i$ in Eq.\ (\ref{7}).
We substitute Eq.\ (\ref{14a}) into Eq.\ (\ref{7}) and divide by $Q_1Q_2\cdots Q_{n+1}$ to obtain the decomposition
\begin{equation}\label{16}
\frac{1}{Q_1Q_2\cdots Q_{n+1}}=\frac{\lambda_1+\mu_1 D_1}{Q_2\cdots Q_{n+1}}+\cdots+\frac{\lambda_{n+1}+\mu_{n+1} D_{n+1}}{Q_1\cdots Q_{n}}
\end{equation}
with
\begin{equation}\label{17}
\mu_i=\frac{(-1)^i \lambda_{n+2}}{\hbox{det}(\bar{p}_1,\dots,\bar{p}_{n+1})}.
\end{equation}
Before we continue we have to implement the result of Step 1, Sect.\ \ref{step1}. To this end it is convenient to use a multi-index notation.
If $I$ is a subset of $I_N=\{1,2,\ldots,N\}$ we write $p_I$ for the sequence of vectors $p_i$, $i\in I$ ordered by the values of their indices,
\begin{equation}\label{25a}
Q^I=\prod_{i\in I}Q_i\quad\hbox{and}\quad D_I=\hbox{det}(\bar{p},\bar{p}_I),
\end{equation}
where the last equation only making sense if $|I|=n$.
Eq.\ (\ref{16}) now states that for any $I_{n+1}$ with $|I_{n+1}|=n+1$ we have
\begin{equation}\label{16a}
\frac{1}{Q^{I_{n+1}}}=\sum_{I\subset I_{n+1}\atop |I|=n}\frac{\lambda_I+\mu_I D_I}{Q^I}.
\end{equation}
The result of Step 1 is a decomposition of $N$ propagators into a sum of products of $n+1$ propagators,
\begin{equation}\label{16b}
\frac{1}{Q^{I_N}}=\sum_{I_{n+1}\subset I_N\atop |I_{n+1}|=n+1}\frac{\nu_{I_{n+1}}}{Q^{I_{n+1}}},
\end{equation}
for some $\nu_{I_{n+1}}\in\R$. Putting both equations together we obtain
\begin{equation}\label{16c}
\frac{1}{Q^{I_N}}=\sum_{I\subset I_N\atop |I|=n}\frac{\lambda_I'+\mu_I' D_I}{Q^I}
\end{equation}
with some new constants $\lambda_I'$, $\mu_I'$. The terms with the $\mu_I'$ vanish by Eq.\ (\ref{12}) upon integrating over $p$.

Now, we use the standard method to calculate the coefficients of a partial fraction decomposition:
Multiplication of Eq.\ (\ref{16c}) by $Q^I$ delivers $\lambda_I'+\mu_I' D_I$
if we substitute for $p$ a simultaneous solution of $Q_i=0$, $i\in I$,
\begin{equation}\label{26}
\lambda_I'+\mu_I' D_I=\left.\frac{1}{Q^{I_N-I}}\right|_{Q_i=0,\,i\in I}.
\end{equation}
To solve the system of equations $Q_i=0$ for $p$ we observe that the differences $Q_i-Q_j$ (keep $j\in I$ fixed) are linear in the coordinates of $p$.
We can hence solve $n-1$ linear equations plus one remaining quadratic equation and expect in general two (complex) solutions. The solutions are best
given in geometrical terms. We again follow the construction of Figs.\ 2, 3, this time for the $n$ vectors $(p_i,\pm m_i)$, $i\in I$.
We plot the $(n-1)$-sphere spanned by these vectors and determine its center $c$ and its radius $r$. The center lies in the $(n-1)$-dimensional hyper-surface
spanned by the tips of the vectors $p_i$. In $\R^n$ this hyper-surface has normal unit-vectors $\pm \sigma$ (no orientation intended). Now, we see that
\begin{equation}\label{27}
p_\pm=c\pm {\rm i}r\sigma
\end{equation}
fulfill the identities $(p_\pm^2-p_i)^2=(c-p_i)^2-r^2=-\,m_i^2$ for $i\in I$. Hence, the $p_\pm$ are the simultaneous solutions of $Q_i=0$.
They are $n$-dimensional vectors with complex entries, $p_\pm\in\C^n$.

If we substitute $p_\pm$ in $D_I$ we immediately see that the real part vanishes because it is given by $n!$ times the $n$-dimensional volume of the simplex spanned by the
tips of the vectors $c,p_I$ (all of which lie in a hyper-plane),
\begin{equation}\label{28}
D_I(p_\pm)\in {\rm i}\R.
\end{equation}
Because $\lambda_I'$ and $\mu_I'$ are real numbers independent of $p$ (they are the solution of a linear system of real equations)
we may solve Eq.\ (\ref{26}) for $\lambda_I'$ by taking the real part
\begin{equation}\label{29}
\lambda_I'={\rm Re}\left.\frac{1}{Q^{I_N-I}}\right|_{p=p_\pm}.
\end{equation}
It makes no difference whether we substitute $p=p_+$ or $p=p_-$ as they are complex conjugates.
We summarize the result of Step 1 and Step 2 in the following formula:
\begin{equation}\label{30}
\int{\rm d}^np\frac{1}{Q_1\cdots Q_N}={\rm Re}\sum_{I\subset \{1,...,N\}\atop |I|=n}
\frac{1}{\prod\limits_{i\in\!\!\!\!/\,I}Q_i}\Bigg|_{p:{Q_i(p)=0\atop i\in I}}
\int{\rm d}^np\,\frac{1}{\prod\limits_{i\in I}Q_i}.
\end{equation}
The two solutions of ${Q_i=0,\,i\in I}$ are given by Eq.\ (\ref{27}). This equation was found in the two-dimensional case ($n=2$) by K\"all\'en and Toll \cite{KT}
and generalized to arbitrary dimensions by Petersson \cite{PE}.
Note that, although we used only elementary methods to prove Eq.\ (\ref{30}), the result resembles a residue formula (think e.g.\ of a complex contour integral over a rational function).
If one replaces the real part by the algebraic trace (the solutions of ${Q_i=0,\,i\in I}$ give a quadratic extension of the field of rational functions
in the momenta and masses and the complex conjugation equals the algebraic conjugation swapping the sign of the square root) the author assumes that
there exists an equivalent formula at any loop order.

\subsection{Step 3: hyperbolic volumes}\label{step3}

The transition to the calculation of hyperbolic volumes is facilitated by the introduction of Feynman parameters.
We need a particularly flexible variant of Feynman parameters that starts from the elementary identity
\begin{equation}\label{31}
\frac{1}{Q_1\cdots Q_N}=(N-1)!\int_0^\infty {\rm d}\alpha_2\cdots\int_0^\infty {\rm d}\alpha_N
\frac{1}{(Q_1+\alpha_2Q_2+\ldots+\alpha_NQ_N)^N}.
\end{equation}
Now, we want to homogenize the denominator by introducing a variable $\alpha_1$. In fact we
move to projective space when we write the integrand as $(\alpha_1Q_1+\ldots\alpha_NQ_N)^{-N}$:
The integrand is a homogeneous (rational) function in the $\alpha$'s of degree $-N$.
To obtain a total degree 0 we multiply the integrand by the $(N-1)$-form $\Omega$ of degree $N$,
\begin{equation}\label{32}
\Omega=\sum_{i=1}^N(-1)^{i-1}\alpha_i{\rm d}\alpha_1\wedge\ldots\wedge{\rm d}\alpha_{i-1}\wedge{\rm d}\alpha_{i+1}\wedge\ldots\wedge{\rm d}\alpha_N.
\end{equation}
If we multiply $\Omega$ by a homogeneous function $f$ of degree $d$ we find that the exterior derivative of the product is given by
\begin{equation}\label{33}
{\rm d}f\Omega=\sum_{i=1}^N((\partial_i f)\alpha_i+f)\,{\rm d}\alpha_1\wedge\ldots\wedge{\rm d}\alpha_N=(d+N)f{\rm d}\alpha_1\wedge\ldots\wedge{\rm d}\alpha_N.
\end{equation}
In our case $f=(N-1)!(\alpha_1Q_1+\ldots+\alpha_NQ_N)^{-N}$ and $f\Omega$ is a closed.
If we integrate $f\Omega$ over the (degenerate) simplex $\alpha_1=$ const $>0$, $\alpha_i \geq 0$, $i=2,\ldots,N$ endowed with
an 'outward' orientation we are back at Eq.\ (\ref{31}).
But, because $f\Omega$ is closed, we are free to deform the sheet of integration as long as we do not pass through the singular origin
$\alpha_1=\ldots=\alpha_N=0$. Another option for the integration surface is the
standard simplex $\alpha_1+\ldots+\alpha_N=1$, $\alpha_i \geq 0$, $i=1,\ldots,N$ which leads to the
standard form of the Feynman parameters. In projective terms we may speak of a (projective) integral in $\Pp^{N-1}\R$ over the
(projective) simplex $\Delta=\alpha_i \geq 0$, $i=1,\ldots,N$. In what follows it will be crucial to have this flexibility in shaping the integration sheet.

Before we continue we need the change of $\Omega$ under linear transformations
\begin{equation}\label{33a}
\alpha_i {\mapsto} \sum_{j=1}^Na_{ij}\beta_j,\quad A=\left(a_{ij}\right)_{i,j}.
\end{equation}
We have
\begin{equation}\label{33b}
\Omega\mapsto \sum_{j_1,\ldots,j_N=1}^Na_{1j_1}\cdots a_{Nj_N}\sum_{i=1}^N(-1)^{i-1}\beta_{j_i}{\rm d}\beta_{j_1}\wedge\ldots
\wedge{\rm d}\beta_{j_{i-1}}\wedge{\rm d}\beta_{j_{i+1}}\wedge\ldots\wedge{\rm d}\beta_{j_N}.
\end{equation}
The sum over $i$ on the right hand side may be written as $(\beta_{j_1}{\rm d}\beta_{j_2}-\beta_{j_2}{\rm d}\beta_{j_1})\wedge\omega_1+
{\rm d}\beta_{j_1}\wedge{\rm d}\beta_{j_2}\wedge\omega_2$ exhibiting an anti-symmetry under exchange of $j_1$ and $j_2$. By symmetry it is
in fact totally anti-symmetric. For $j_1=1,\ldots,j_N=N$ we are back at the definition (Eq.\ (\ref{32})) of $\Omega$. The sum over $i$ is
thus $\epsilon_{j_1\ldots j_N}\Omega$ (where $\epsilon$ is the totally anti-symmetric tensor) and we obtain
\begin{equation}\label{33c}
\Omega\mapsto \hbox{det}(A)\,\Omega.
\end{equation}

Now, we proceed as usual and interchange the $p$- and the $\alpha$-integrals. After a shift $p\mapsto p+(\sum\alpha_ip_i)/(\sum\alpha_i)$ the
$p$-integral has rotational symmetry and we obtain after the trivial angular integral is evaluated (we do not yet use Eq.\ (\ref{30}))
\begin{equation}\label{34}
\int{\rm d}^np\frac{1}{Q_1\cdots Q_N}=\int_\Delta\Omega\,\frac{2\pi^{\frac{n}{2}}}{\Gamma(\frac{n}{2})}\int_0^\infty\frac{{\rm d}p\,p^{n-1}}
{\left((\sum\alpha_i)p^2+\frac{(\sum\alpha_i)(\sum\alpha_i(p_i^2+m_i^2))-(\sum\alpha_ip_i)^2}{\sum\alpha_i}\right)^N}.
\end{equation}
The (one-dimensional) $p$-integral leads to the beta-function with the result
\begin{equation}\label{35}
\int\frac{{\rm d}^np}{Q_1\cdots Q_N}=\pi^{\frac{n}{2}}\Gamma\!\left(N-\frac{n}{2}\right)\!\int_\Delta\!\!\Omega\left(\sum_{i=1}^N\alpha_i\right)^{\!N-n}
\!\left(\sum_{i,j=1}^N\alpha_i\alpha_j\frac{(p_i-p_j)^2+m_i^2+m_j^2}{2}\right)^{\!\frac{n}{2}-N}.
\end{equation}
There are two cases where the integrand simplifies: One, $N=n/2$ is the log-divergent case. It demands regularization which partly spoils its simplicity.
We will not pursue this case here. Two, $N=n$ where the $\sum\alpha_i$-term is absent. This case is convergent and we can reduce the case $N>n$ to this case by
virtue of steps 1 and 2, Eq.\ (\ref{30}).
 
 For $N=n$ we may integrate over the surface
\begin{equation}\label{36}
\Delta_1=\left\{\alpha_i:\sum_{i,j=1}^n\alpha_i\alpha_j\frac{(p_i-p_j)^2+m_i^2+m_j^2}{2}=1,\; \alpha_i\geq 0\right\}
\end{equation}
and we obtain
\begin{equation}\label{37}
\int{\rm d}^np\frac{1}{Q_1\cdots Q_n}=\pi^{\frac{n}{2}}\Gamma\left(\frac{n}{2}\right)\int_{\Delta_1}\Omega.
\end{equation}
Specifying the integration surface we lose the projective property of the integral and we have to interpret the right hand side as a
standard $(n-1)$-dimensional surface integral. One the other hand, the right hand side has now the shape of an area integral.
To actually convert it into an area integral we perform a linear transformation
\begin{equation}\label{38}
u^{(j)}=\sum_{i=1}^n\alpha_iP_i^{(j)},\quad j=1,\ldots,n,
\end{equation}
where the $P_i$ form an $n$-tuple of vectors in $\C^n$. If we choose
\begin{equation}\label{39}
P_i=p_i-p_\pm
\end{equation}
with $p_\pm$ either $p_+$ or $p_-$ from Eq.\ (\ref{27}) we find for their scalar products
\begin{eqnarray}\label{40}
P_i\cdot P_j&=&(p_i-c)\cdot(p_j-c)-r^2=\frac{(p_i-c)^2+(p_j-c)^2-(p_i-p_j)^2}{2}-r^2\nonumber\\
&=&-\frac{(p_i-p_j)^2+m_i^2+m_j^2}{2}.
\end{eqnarray}
In particular
\begin{equation}\label{40a}
P_i^2=-m_i^2.
\end{equation}
We may thus interpret $\Delta_1$ in terms of our new variables $u^{(j)}$, $j=1,\ldots,n$, as
\begin{equation}\label{41}
\Delta_1=\{u:u^2=-1\}\cap \hbox{cone}_{\R_+}(P_1,...,P_n).
\end{equation}

\begin{figure}[ht]
\epsfig{file=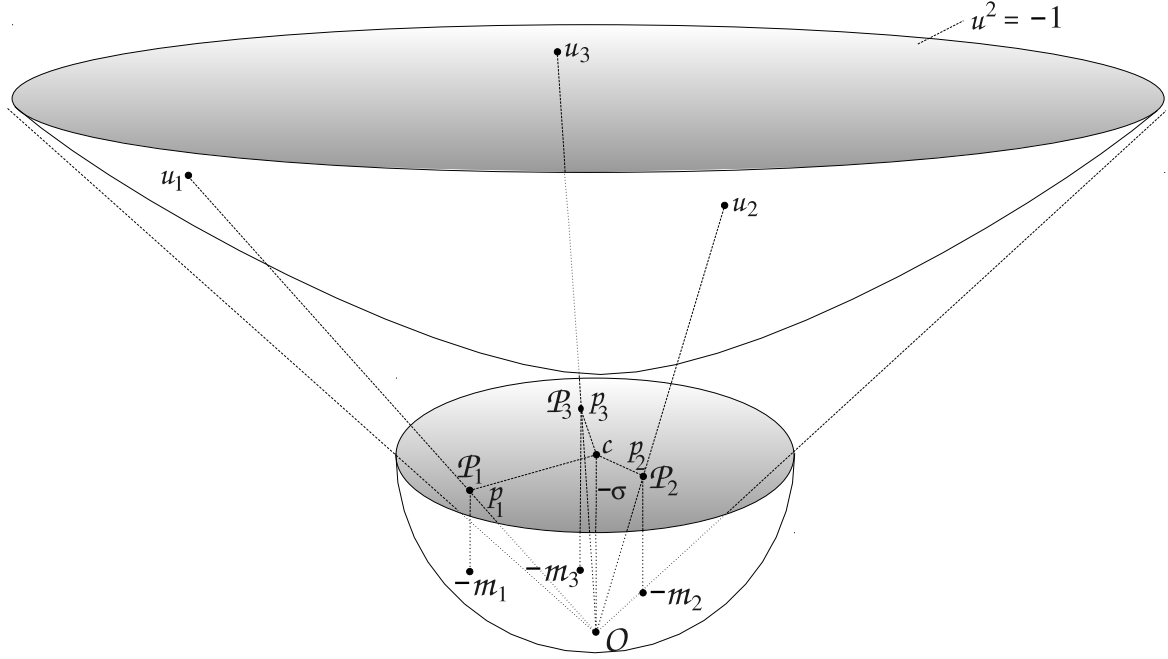,width=\textwidth}
\caption{Hyperbolic picture of the one-loop amplitude in Minkowski 3-space. Note that we start in $n=3$ euclidean dimensions
in which the tips of the three vectors $p_1$,$p_2$,$p_3$ span a surface. Around this plane we construct a three-dimensional Minkowski space
by introducing a vertical 'mass'-dimension of negative signature. Here, we suppress the transverse dimension of the original euclidean space
to obtain a three-dimensional picture.}
\end{figure}

Here $u\in\hbox{cone}_{\R_+}(P_1,...P_n)$ means nothing but Eq.\ (\ref{38}) for non-negative (real) $\alpha_i$.
Now we have a nice description of $\Delta_1$ as a real subset in a complex vector space. In order to obtain a
more conventional picture we want to go back to a real vector space. We can do so by absorbing the i in
Eq.\ (\ref{27}) into the metric thus passing to Minkowski space with signature $(-,+,\ldots,+)$.
The origin of Minkowski space is located at $c-\sigma r$ (or at $c+\sigma r$), see Fig.\ 4. The vectors
$p_i-c$ lie in the ($n-1$)-dimensional euclidean subspace with origin $c=(r,0)$ and point from $c$ to the
vectors $P_i$ which become Minkowski $n$-momenta with coordinates\footnote{Strictly speaking the Minkowski coordinates
are only defined up to an arbitrary rotation of the ($n-1$)-dimensional euclidean subspace. We use (any set of) coordinates in
the intermediate steps; the results are expressed in terms of Minkowski invariants.}
\begin{equation}\label{42}
P_i^{\rm M}=(r,p_i^{\rm M}),\quad (P_i^{\rm M})^2=-m_i^2.
\end{equation}
Equations (\ref{40}) to (\ref{41}) remain valid for Minkowski space vectors. We thus may interpret $\Delta_1$ as
\begin{equation}\label{43}
\Delta_1=\hbox{Set of $n$-velocities spanned by }u_i^{\rm M}=P_i^{\rm M}/m_i.
\end{equation}
A standard calculation shows that the Minkowski metric $g=-{\rm d}u_0^2+{\rm d}u_1^2+\ldots+{\rm d}u_{n-1}^2$
induces on the surface of $n$-velocities $-u_0^2+u_1^2+\ldots+u_{n-1}^2=-1$ the measure $\Sigma_1=1/u_0\cdot{\rm d}u_1\wedge\ldots\wedge{\rm d}u_{n-1}$
(eliminate ${\rm d}u_0=\sum u_i{\rm d}u_i/u_0$). We obtain the same measure from the projective form $\Omega$ in the $u^{\rm M}$ variables.
The transformation Eq.\ (\ref{38}) from the $\alpha$-variables to the $u$-variables gives a 'Jacobian' by Eq.\ (\ref{33c}). Altogether we obtain
\begin{equation}\label{44}
\Omega=\frac{1}{\hbox{det}(P_1^{\rm M},\ldots,P_n^{\rm M})}\Sigma_1\quad\hbox{on }\Delta_1.
\end{equation}
On the other hand the Minkowski metric induces on the surface of $n$-velocities ('sphere of radius i') a hyperbolic space of constant
curvature $-1$. The cone spanned by the $P_i^{\rm M}$ cuts out a simplex spanned by the $u_i^{\rm M}$ with geodesic faces: a hyperbolic (hyper-)tetrahedron $\Sigma$
(see e.g.\ \cite{TH}).

The determinant in the denominator of Eq.\ (\ref{44}) is readily calculated:
\begin{eqnarray}\label{45}
-\hbox{det}(P_1^{\rm M},\ldots,P_n^{\rm M})^2&=&\hbox{det}\left[(P_1^{\rm M},\ldots,P_n^{\rm M})^T\cdot\hbox{diag}(-1,1,\ldots,1)\cdot(P_1^{\rm M},\ldots,P_n^{\rm M})\right]\\\nonumber
&=&\hbox{det}\left[(P_i^{\rm M}\cdot P_j^{\rm M})_{i,j}\right]\\\nonumber
&=&\hbox{det}\left[\left(-\frac{(p_i-p_j)^2+m_i^2+m_j^2}{2}\right)_{i,j}\right]
\end{eqnarray}
by Eq.\ (\ref{40}). Now, we can express the one-loop amplitude in terms of the volume of the ($n-1$)-dimensional hyperbolic simplex $\Sigma$ spanned by $u_1^{\rm M},\ldots,u_n^{\rm M}$
\begin{equation}\label{46}
\int{\rm d}^np\frac{1}{Q_1\cdots Q_n}=\frac{(2\pi)^\frac{n}{2}\Gamma\left(\frac{n}{2}\right)\hbox{vol}_{H^{n-1}}[\Sigma(u_1^{\rm M},\ldots,u_n^{\rm M})]}
{[(-1)^{n-1}\hbox{det}((p_i-p_j)^2+m_i^2+m_j^2)_{i,j}]^{1/2}}.
\end{equation}
Note that the absolute of det$(P_1^{\rm M},\ldots,P_n^{\rm M})$ is given by $n!$ times the volume
of the simplex spanned by the $P_i^{\rm M}$ or, by Eq.\ (\ref{42}), $(n-1)!$ times the radius $r$ times the area of the surface spanned by the tips of
the vectors $p_i$. We can use a Minkowski version of Eq.\ (\ref{11}) for this case and solve for $r$ to obtain (compare Eq.\ (\ref{4}) and Fig.\ 2 for the case $n=2$)
\begin{equation}\label{47}
r=\left[\frac{(-1)^{n-1}\hbox{det}((p_i-p_j)^2+m_i^2+m_j^2)_{i,j=1\ldots n}}{2^n\,\hbox{det}((p_i-p_1)\cdot(p_j-p_1))_{i,j=2\ldots n}}\right]^{1/2}.
\end{equation}
The coordinates of the center $c$ can be calculated by linear algebra. Because $c$ is in the hyperplane spanned by the tips of $p_i$ we have
\begin{equation}\label{48}
c=\sum_{i=1}^n\lambda_ip_i\,,\quad \sum_{i=1}^n\lambda_i=1
\end{equation}
and from Eq.\ (\ref{23}) we get the system of linear equations
\begin{equation}\label{49}
\sum_{i=1}^n\lambda_ip_i\cdot(p_j-p_1)=\frac{p_j^2-p_1^2+m_j^2-m_1^2}{2},\quad j=2,\ldots,n.
\end{equation}
The previous two equations can be summarized to the matrix equation
\begin{equation}\label{50}
\left(\begin{array}{ccc}
1&\ldots&1\\
p_1\cdot(p_2-p_1)&\ldots&p_n\cdot(p_2-p_1)\\
\vdots&&\vdots\\
p_1\cdot(p_n-p_1)&\ldots&p_n\cdot(p_n-p_1)\end{array}\right)\cdot\lambda=
\left(\begin{array}{c}
1\\
\frac{p_2^2-p_1^2+m_2^2-m_1^2}{2}\\
\vdots\\
\frac{p_n^2-p_1^2+m_n^2-m_1^2}{2}\end{array}\right)
\end{equation}
which is solved by
\begin{equation}\label{51}
\lambda_i=\frac{
\raisebox{-1ex}{\hbox{det}}
\begin{array}{c}i\\
\left(\begin{array}{ccccc}
1&\ldots&1&\ldots&1\\
p_1\cdot(p_2-p_1)&\ldots&\frac{p_2^2-p_1^2+m_2^2-m_1^2}{2}&\ldots&p_n\cdot(p_2-p_1)\\
\vdots&&\vdots&&\vdots\\
p_1\cdot(p_n-p_1)&\ldots&\frac{p_n^2-p_1^2+m_n^2-m_1^2}{2}&\ldots&p_n\cdot(p_n-p_1)\end{array}\right)\end{array}}
{\hbox{det}((p_i-p_1)\cdot(p_j-p_1))_{i,j=2\ldots n}}.
\end{equation}

\begin{figure}[ht]
\hspace{5cm}\epsfig{file=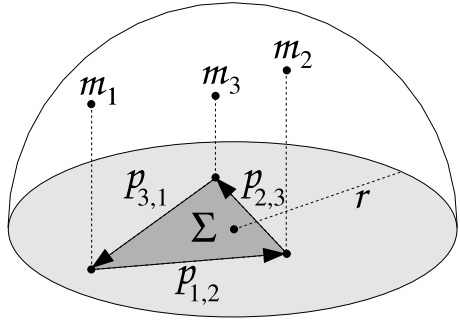,width=6cm}
\caption{The geometry of the one-loop amplitude in $n=3$ dimensions.}
\end{figure}

We end this subsection by formulating the result in terms of the external ingoing momenta $p_{i,i+1}=p_{i+1}-p_i$ using the projective model
of hyperbolic space. In this model geodesics are straight lines. Because the sum of the external momenta is zero we obtain a simplex $\Sigma$ as
convex hull of the vector sum of the external momenta (see Fig.\ 5). If we interpret this simplex as a tetrahedron in projective hyperbolic
space with the 'sphere at infinity' given by the construction of Fig.\ 3 we can formulate Eq.\ (\ref{46}) purely in terms of geometrical quantities
\begin{equation}\label{52}
\int{\rm d}^np\frac{1}{Q_1\cdots Q_n}=\frac{\hbox{vol}(S^n_{1/2})\hbox{vol}_{H^{n-1}}(\Sigma)}{r\,\hbox{vol}_{\R^{n-1}}(\Sigma)}.
\end{equation}
Here, $S^n_{1/2}$ is the $n$-sphere of radius $1/2$. In order to use Eqs.\ (\ref{47}) and (\ref{51}) to calculate $r$ and the origin of the projective
model we may set $p_1=0$ and $p_i=\sum_{j=1}^{i-1}p_{j,j+1}$.

Note that a (simpler) result for $N=n-1$ external momenta (if convergent) can be derived from Eq.\ (\ref{46}) by taking an appropriate limit $p_n\mapsto \infty$.
\pagebreak[3]

{\bf Example 1: }$n=2$.

Eqs.\ (\ref{40}), (\ref{40a}) determine the length of the line $u_1^{\rm M}u_2^{\rm M}$ (see \cite{TH}) and Eq.\ (\ref{46}) yields \cite{KT}
\begin{equation}\label{53a}
\int{\rm d}^2p\frac{1}{Q_1Q_2}=\frac{2\pi\,\hbox{arcosh}\left(\frac{p_{1,2}^2+m_1^2+m_2^2}{2m_1m_2}\right)}{([p_{1,2}^2+(m_1+m_2)^2][p_{1,2}^2+(m_1-m_2)^2])^{1/2}}.
\end{equation}
{\bf Example 2: }$n=3$.

In $n=3$ dimensions we have to calculate the hyperbolic area of a triangle in two-dimensional hyperbolic space.
The result $\pi$ minus the sum of the interior angles (see Eq.\ (\ref{58})) is a special case of the Gau{\ss}-Bonnet formula reducing
volumes of tetrahedra in even-dimensional hyperbolic space to objects of lower dimension. This will be pursued in the following subsection.

Here, we first specialize to the massless case $m_i=0$. In this situation the triangle is ideal having its vertices at infinity.
The interior angles are all zero (note that the projective model does not present angles faithfully) leading to a hyperbolic area $\pi$ independent
of the momenta (in hyperbolic space polyhedra have finite volume even if their vertices are at infinity). Eq.\ (\ref{46}) hence yields \cite{NI}
\begin{equation}\label{53}
\left.\int{\rm d}^3p\frac{1}{Q_1Q_2Q_3}\right|_{m_i=0}=\frac{\sqrt{2}\pi^2\cdot\pi}{[\hbox{det}(|p_i-p_j|^2)_{i,j=1\ldots3}]^{1/2}}
=\frac{\pi^3}{|p_1-p_2||p_2-p_3||p_3-p_1|}.
\end{equation}

For general masses we have Eqs.\ (\ref{40}), (\ref{40a}) to determine the lengths of the sides of the triangle (use the hyperboloid model \cite{TH})
\begin{equation}\label{54}
\cosh(d_{i,j})=\frac{|p_i-p_j|^2+m_i^2+m_j^2}{2m_im_j}
\end{equation}
and the triangle formula ($i$, $j$, $k$, are mutually different)
\begin{equation}\label{55}
\cos(\alpha_i)=\frac{\cosh(d_{i,j})\cosh(d_{i,k})-\cosh(d_{j,k})}{\sinh(d_{i,j})\sinh(d_{i,k})}
\end{equation}
to calculate the interior angles $\alpha_i$. Note, that the result amounts to taking square roots and inverting the cosine giving rise (in $\C$) to
a logarithm which was already present at $n=2$ dimensions (Eq.\ (\ref{53a})). This kind of reduction from odd to even dimensions will be studied in the next subsection.

\subsection{Step 4: $n$ odd $\rightarrow n-1$ even}

This step is only needed in case one is interested in odd-dimensional space-time. It basically shows that
one-loop calculations in odd $n$ dimensions are not significantly harder than in $n-1$ dimensions. Except
for algebraic functions (like square roots) one can use the same set of functions to express the result
in $n=2m+1$ dimensions as in $n\leq 2m$ dimensions: In particular, in 3 dimensions the one-loop amplitude
can be expressed in terms of logs (see Ex.\ 2) and in 5 dimensions in terms of dilogarithms and logarithms.

The reduction is facilitated by a generalized version of the classical Gau{\ss}-Bonnet formula (\cite{BE}, Th\'eor\`eme 1.2):
The Euler-characteristic of a tetrahedron is given by a sum over volumes times normal dihedral angles of even-dimensional subspaces.
To be specific, for an $n$-dimensional tetrahedron in a space of constant curvature $\kappa$ we have
\begin{equation}\label{56}
\chi=\sum_{i=0,2,4,\ldots}^n\frac{2\kappa^{i/2}}{\hbox{vol}(S^i)\hbox{vol}(S^{n-i-1})}\sum_{i{\rm -faces}\,\sigma}\hbox{vol}(\sigma)\cdot \hbox{normal-dihedral-angle}(\sigma)
\end{equation}

In hyperbolic space we have $\kappa=-1$ and the normal dihedral angle is the volume of the spherical angle of the cone dual to the cone of the dihedral angle.
By definition the normal dihedral angle of the full polygon, the volume of a point, and vol($S^{-1}$) are 1. The Euler-characteristic of a polygon is one and for even dimension $n=2m$ we can
solve for the hyperbolic volume of the tetrahedron and obtain
\begin{equation}\label{57}
\hbox{vol}(\Sigma)=\frac{(-1)^m\pi^{m+\frac{1}{2}}}{\Gamma(m+\frac{1}{2})}-\sum_{i=0}^{m-1}\frac{(-1)^{m-i}\beta(i+\frac{1}{2},m-i)}{2}
\!\!\!\sum_{2i{\rm -faces}\,\sigma}\!\!\!\!\hbox{vol}(\sigma)\cdot \hbox{normal-dihedral-angle}(\sigma)
\end{equation}
where $\beta(x,y)=\Gamma(x)\Gamma(y)/\Gamma(x+y)$ is the beta-function.

In two dimensions ($m=1$) the normal dihedral angles are $\pi$ minus the interior angles and we find the well-known formula
\begin{equation}\label{58}
\hbox{vol}(\Sigma)=-2\pi+\sum_{k=1}^3(\pi-\alpha_k)=\pi-\alpha_1-\alpha_2-\alpha_3.
\end{equation}

In four dimensions ($m=2$) we have
\begin{eqnarray}\label{59}
\hbox{vol}(\Sigma)&=&\frac{4\pi^2}{3}-\frac{2}{3}\sum_{{\rm vertices}\,x}\hbox{normal-dihedral-3-angle}\,(x)\nonumber\\
&&+\frac{1}{3}\sum_{2{\rm -faces}\,\sigma}(\pi-\sum\hbox{interior-angles}(\sigma))(\pi-\hbox{dihedral-angle}(\sigma))
\end{eqnarray}
which simplifies in the case of an ideal tetrahedron to
\begin{equation}\label{59a}
\hbox{vol}(\Sigma)=\frac{\pi}{3}\left(4\pi-\sum_{2{\rm -faces}\,\sigma}\hbox{dihedral-angle}(\sigma)\right).
\end{equation}

In general, repeated use of Eq.\ (\ref{57}) reduces the evaluation of even-dimensional hyperbolic volumes to
the calculation of odd-dimensional normal dihedral angles. These angles are spherical simplices and calculations in spherical
geometry are in some sense dual to calculations in hyperbolic geometry. By suitable changes of signs and the replacement of trigonometric
functions by their hyperbolic counterparts we can often express spherical quantities by their hyperbolic analogons.
For example, in spherical 2-space the area of a triangle is given by $\alpha_1+\alpha_2+\alpha_3-\pi$ in complete analogy to Eq.\ (\ref{58}).
Although we keep the general correspondence somewhat vague, we restrict ourselves in the last step to odd-dimensional hyperbolic space (corresponding to even-dimensional space-time).

\subsection{Step 5: $m_i=0$, ideal tetrahedra}\label{step5}

\begin{figure}[ht]
\hspace{5cm}\epsfig{file=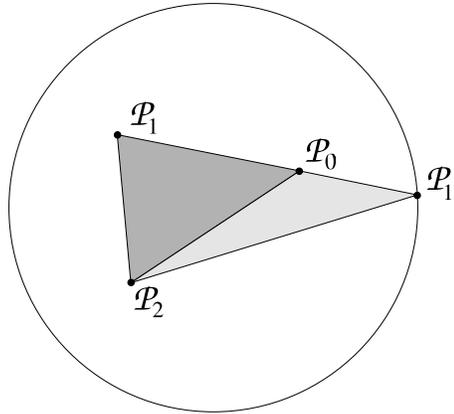,width=6cm}
\caption{Decomposition of simplices into ideal tetrahedra, step one: moving all but one vertex to infinity. The triangle $P_0P_1P_2$ can be expressed as
the triangle $P_{1'}P_1P_2$ minus the triangle $P_0P_{1'}P_2$. Next, one proceeds with the ray $P_2P_1$ or $P_2P_0$ (respectively) to decompose the
triangle $P_0P_1P_2$ into an alternating sum of four triangles each of which has two vertices at infinity.}
\end{figure}

For any simplex with vertices $P_0,\ldots,P_n$ we have the option to choose a point $P_{1'}=\lambda P_0+(1-\lambda) P_1$, $\lambda>1$ on the ray $P_1P_0$ (say)
outside of the simplex and express the simplex as a difference of the simplices $P_{1'}P_1P_2\ldots P_n$ minus $P_{1'}P_0P_2\ldots P_n$.
In the projective model this means (see Fig.\ 6)
\begin{equation}\label{60}
\hbox{vol}(P_0,\ldots,P_n)=\hbox{vol}(P_{1'},P_1,P_2,\ldots,P_n)-\hbox{vol}(P_0,P_{1'},P_2,\ldots,P_n).
\end{equation}

We may choose $P_{1'}$ at infinity (as the intersection of the ray $P_1P_0$ with the sphere at infinity) and if neither $P_0$ nor $P_1$ was at infinity
the simplices on the right hand side have one additional ideal vertex. We can proceed until all but one vertex are at infinity:
A general simplex is the alternating sum of $2^n$ simplices with all but one vertex at infinity.

By construction (see Fig.\ 5) vertices at infinity translate into massless propagators. The above construction is thus precisely the geometric version
of the decomposition of Step 0, Sect.\ \ref{step0}.

In odd-dimensional hyperbolic space we can do more. A simplex with all but one vertex at infinity can be expressed as an alternating
sum of ideal tetrahedra (with all vertices at infinity). This peculiar property is very special to odd-dimensional hyperbolic space. It is given by the following construction
(see Fig.\ 7):

\begin{figure}[ht]
\hspace{5cm}\epsfig{file=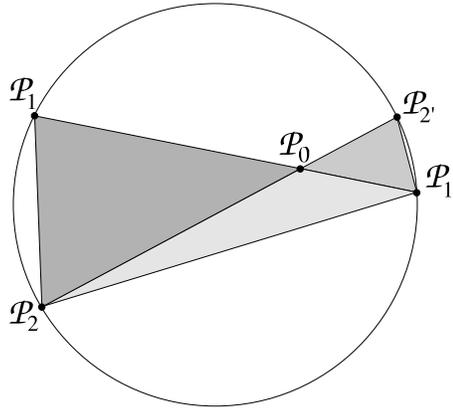,width=6cm}
\caption{Decomposition of simplices into ideal tetrahedra, step two: moving the last vertex to infinity. The triangle $P_0P_1P_2$ can be expressed as
the triangle $P_{1'}P_1P_2$ minus the triangle $P_{1'}P_{2'}P_2$ plus the triangle $P_0P_{1'}P_{2'}$. The triangles $P_0P_1P_2$ and $P_0P_{1'}P_{2'}$ map
onto each other by an isometric reflection at $P_0$. In odd dimensions one can solve for the volume of the triangle $P_0P_1P_2$.}
\end{figure}

Let $P_0$ be the only vertex not at infinity and $P_{i'}$ be the intersection of the ray $P_iP_0$ with the sphere at infinity.
Using Eq.\ (\ref{60}) for $P_1=P_k$, $k=1,\ldots,n$ and we obtain
\begin{equation}\label{61}
\hbox{vol}(P_0,P_{1'}\ldots P_{k-1'},P_k\ldots P_n)=\hbox{vol}(P_{1'}\ldots P_{k'},P_k\ldots P_n)-\hbox{vol}(P_0,P_{1'}\ldots P_{k'},P_{k+1}\ldots P_n).
\end{equation}
Taking the alternating sum of the above equations we obtain
\begin{equation}\label{62}
\hbox{vol}(P_0,\ldots,P_n)=\sum_{k=1}^n(-1)^{k-1}\hbox{vol}(P_{1'},\ldots,P_{k'},P_k,\ldots,P_n)+(-1)^n\hbox{vol}(P_0,P_{1'},\ldots,P_{n'}).
\end{equation}
The last simplex on the right hand side may be reflected at $P_0$ without changing its volume yielding the simplex $(P_0,P_1,\ldots,P_n)$.
In odd dimensions we can solve for $\hbox{vol}(P_0,\ldots,P_n)$ to obtain
\begin{equation}\label{63}
\hbox{vol}(P_0,\ldots,P_n)=\frac{1}{2}\sum_{k=1}^n(-1)^{k-1}\hbox{vol}(P_{1'},\ldots,P_{k'},P_k,\ldots,P_n).
\end{equation}
Altogether, in odd-dimensional hyperbolic space we can express any simplex as $1/2$ times the alternating sum of $n2^n$ ideal tetrahedra.
In fact, if one chooses the points at infinity in a clever way many terms drop. At three dimensions one is left with
a sum over 8 ideal tetrahedra, namely e.g.
\begin{eqnarray}\label{64}
\hbox{vol}(P_0,P_1,P_2,P_3)&=&\frac{1}{2}\Big[\hbox{vol}(P_{01},P_{02},P_{23},P_{30})+\hbox{vol}(P_{01},P_{12},P_{32},P_{13})\nonumber\\
&&+\,\hbox{vol}(P_{01},P_{21},P_{23},P_{13})+\hbox{vol}(P_{01},P_{21},P_{32},P_{31})\nonumber\\
&&-\,\hbox{vol}(P_{01},P_{12},P_{23},P_{31})-\hbox{vol}(P_{01},P_{02},P_{32},P_{03})\\
&&-\,\hbox{vol}(P_{01},P_{20},P_{23},P_{03})-\hbox{vol}(P_{01},P_{20},P_{32},P_{30})\Big].\nonumber
\end{eqnarray}
where $P_{ij}$ is the intersection of the ray $P_iP_j$ with the sphere at infinity.

In three dimensions ($n=4$) one can express the hyperbolic volume of ideal tetrahedra in terms of the Bloch-Wigner dilogarithm, Eq.\ (\ref{0f}) (see \cite{BL}, \cite{TH}, \cite{ZA}).
The best way to do this is to interpret the incoming 4-momenta $p_{1,2}$, $p_{2,3}$, $p_{3,4}$, $p_{4,1}$ (which span the ideal tetrahedron
$\Sigma$ in the sense of Fig.\ 5) as quaternions as we did in Sect.\ \ref{4dex}. We have (see Eq.\ (\ref{0g}))
\begin{equation}\label{66}
\hbox{vol}_{H^3}(\Sigma)=|D(p_{1,2}\,p_{2,3}^{-1}\,p_{3,4}\,p_{4,1}^{-1})|.
\end{equation}

In five hyperbolic dimensions ($n=6$ space-time dimensions) one additionally needs a trilogarithm to express the volumes of ideal hyperbolic tetrahedra \cite{KE1}, \cite{KE2}.
Beyond five dimensions not much is known.

\section{A two-loop counter-example}

\begin{figure}[ht]
\hspace{3cm}\epsfig{file=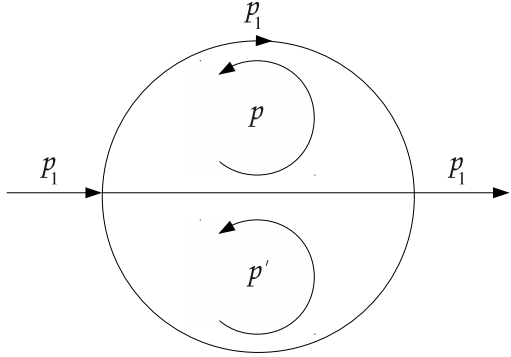,width=10cm}
\caption{A two-loop graph with two external momenta $p_1$ and $-p_1$.}
\end{figure}

Even in the simples possible setup, $n=2$ dimensions, $N=3$ propagators, one obtains an elliptic integral
for the amplitude if all masses are non-zero. This is readily seen by introducing Feynman parameters which yield in the above case
\begin{eqnarray}\label{67}
A(p_1,m_1,m_2,m_3)&=&\int{\rm d}^2\,p\;{\rm d}^2p'\frac{1}{[(p_1-p)^2+m_1^2][(p-p')^2+m_2^2][(p')^2+m_3^2]}\\
&=&\pi^2\int\Omega\frac{1}
{(\alpha_1\alpha_2+\alpha_2\alpha_3+\alpha_3\alpha_1)(\alpha_1m_1^2+\alpha_2m_2^2+\alpha_3m_3^2)+\alpha_1\alpha_2\alpha_3p_1^2}.\nonumber
\end{eqnarray}
If we dehomogenize the integral by setting $\alpha_1=1$ (see Sect.\ \ref{step3}) and explicitly integrate over $\alpha_2$ we obtain
\begin{equation}\label{68}
A\;=\;\frac{\ln\frac{\displaystyle B+\sqrt{B^2-4C}}{\displaystyle B-\sqrt{B^2-4C}}}{\sqrt{B^2-4C}}
\quad\hbox{with}\quad
\begin{array}{l}
B=(1+\alpha_3)(m_1^2+\alpha_3m_3^2)+\alpha_3(m_2^2+p_1^2),\\
\\
C=\alpha_3m_2^2(1+\alpha_3)(m_1^2+\alpha_3m_3^2).
\end{array}
\end{equation}
The denominator is the root of a degree 4 polynomial making the integral the elliptic analogon of the dilogarithm.
Only if one of the masses or the momentum $p_1$ is zero the integral remains rational and is expressible in terms of dilogarithms and logarithms.

\end{document}